\documentclass[conference]{IEEEtran}

\IEEEoverridecommandlockouts

\usepackage{cite}
\usepackage{adjustbox}
\usepackage{subcaption}
\usepackage{comment}
\usepackage{amsmath,amssymb,amsfonts}
\usepackage{algorithmic}
\usepackage{graphicx}
\usepackage{textcomp}
\usepackage{enumitem}
\usepackage{xcolor}
\usepackage{xspace}
\usepackage{array}
\usepackage{makecell}
\usepackage[table]{xcolor} %
\def\BibTeX{{\rm B\kern-.05em{\sc i\kern-.025em b}\kern-.08em
    T\kern-.1667em\lower.7ex\hbox{E}\kern-.125emX}}

\newcommand{\removespacehere}[0]{\vspace{-0.5cm}}

\begin{document}

\title{Taming GPU Underutilization via Static Partitioning and Fine-grained CPU Offloading}

\author{
\IEEEauthorblockN{Gabin Schieffer, Ruimin Shi}
\IEEEauthorblockA{\textit{KTH Royal Institute of Technology}\\
Stockholm, Sweden \\
\{gabins, ruimins\}@kth.se}
\and
\IEEEauthorblockN{Jie Ren}
\IEEEauthorblockA{\textit{William \& Mary}\\
Williamsburg, USA \\
jren03@wm.edu}
\and
\IEEEauthorblockN{Ivy Peng}
\IEEEauthorblockA{\textit{KTH Royal Institute of Technology}\\
Stockholm, Sweden \\
ivybopeng@kth.se}
}

\pagestyle{plain}

\maketitle

\begin{abstract}

Advances in GPU compute throughput and memory capacity brings significant opportunities to a wide range of workloads. However, efficiently utilizing these resources remains challenging, particularly because diverse application characteristics may result in imbalanced utilization. Multi-Instance GPU (MIG) is a promising approach to improve utilization by partitioning GPU compute and memory resources into fixed-size slices with isolation. Yet, its effectiveness and limitations in supporting HPC workloads remain an open question. We present a comprehensive system-level characterization of different GPU sharing options using real-world scientific, AI, and data analytics applications, including NekRS, LAMMPS, Llama3, and Qiskit. Our analysis reveals that while GPU sharing via MIG can significantly reduce resource underutilization, and enable system-level improvements in throughput and energy, interference still occurs through shared resources, such as power throttling. Our performance-resource scaling results indicate that coarse-grained provisioning for tightly coupled compute and memory resources often mismatches application needs. To address this mismatch, we propose a memory-offloading scheme that leverages the cache-coherent Nvlink-C2C interconnect to bridge the gap between coarse-grained resource slices and reduce resource underutilization.

\end{abstract}


\section{Introduction}
The rapid growth of machine learning, scientific, and data analytics workloads has driven an unprecedented demand for GPU resources. For example, Large Language Models (LLM) require high memory capacity, bandwidth, and compute throughput to meet the performance target. On recent generations of GPU architectures, compute and memory resources are continuously increasing. For instance, in the latest four generations of Nvidia GPUs, GPU compute and memory capacity are evolving at a nearly doubling rate as summarized in Table~\ref{tab:mem_compute_gpus}. While the increased GPU resources can generally benefit a wide range of workloads, the exact performance-resource scaling may differ in applications, depending on their characteristics. Consequently, their utilization of GPU resources may become suboptimal due to workload-specific characteristics in resource demands, raising the concern of under-utilization of GPU resources, which in turn undermines overall operating cost-efficiency of GPU systems~\cite{han2022microsecond,strati2024orion}.

GPU sharing emerges as a promising approach to improve resource utilization by enabling multiple processes to concurrently run a single GPU~\cite{gregg2012fine,han2022microsecond,li2022miso,li2022characterizing,wu2023transparent,weaver2024granularity}. Two main GPU sharing schemes exist -- \textit{temporal sharing} and \textit{spatial sharing}. Time slicing on Nvidia GPUs, for instance executes one process at a time on the GPU, switching among processes to create the illusion of multiple processes making progress at the same time. Spatial sharing partitions compute and memory resources among processes so that multiple processes can execute on the GPU concurrently. Common spatial sharing mechanisms include \textit{Multi-Process Service}~(MPS)~\cite{nvidia_mps}, which partitions compute resources only, and the more recent \textit{Multi-Instance GPU}~(MIG), which partitions both compute and memory resources. These GPU sharing options provide different levels of isolation, flexibility, and resource overhead~\cite{pavlidakis2024guardian}. Many prior works have explored GPU sharing for machine learning workloads~\cite{shi2025nexus,tan2021serving,li2022characterizing}, but few have explored resource underutilization in HPC applications with a particular focus on the MIG mechanism~\cite{gilman2022characterizing,weaver2024granularity}.

\begin{table}[bt]
    \centering
    \caption{Characteristics of four generations of Nvidia GPUs.}
    \label{tab:mem_compute_gpus}
    \begin{tabular}{|c|c|c|c|c|c|}
        \hline
              & \multicolumn{2}{c|}{Memory} & \multicolumn{3}{c|}{Throughput (TFLOPS)} \\
              & Capacity & Bandwidth & FP32 & Tensor (FP16) & SMs\\
        \hline\hline
         V100 & 32~GB    & 1.1~TB/s & 16.4 & 130 & 80 \\ \hline
         A100 & 80~GB    & 2.0~TB/s & 19.5 & 312 & 108\\ \hline
         H100 & 144~GB   & 4.9~TB/s & 60   & 1000 & 132\\ \hline
         B200 & 192~GB   & 8.0~TB/s & 80   & 2500 & 160 \\ \hline
    \end{tabular}
    \vspace{-0.3cm}
\end{table}

In this work, we focus on the impact of GPU sharing on resource underutilization, both regarding compute resources, in terms of the number of streaming multiprocessors (SMs), and memory resources, in terms of capacity and bandwidth. We compose a suite of eight real-world workloads, including Qiskit, LAMMPS, NekRS, Llama~3, and FAISS, to represent workloads in scientific, LLM, and data analytics domains. On a Nvidia Grace Hopper system, we collect a set of metrics pertaining to compute and memory resource underutilization, to assess the effectiveness of different GPU sharing options, such as MPS and MIG, in improving resource efficiency. Our results indicate that the coarse-grained resource provisioning of tightly coupled compute and memory resources in MIG profiles mismatches the resource-performance scaling characteristics in real applications.

To address the coarse-grained resource provisioning in fixed MIG slices, we further explore leveraging Nvlink-C2C to selectively offload memory onto CPU, so that workloads slightly larger than a MIG slice can run without requiring a larger MIG slice. Since MIG slices increase compute and memory resources at a doubling rate, this offloading option improves the continuity in resource provisioning. Consequently, this approach can be used to address the problem of resource underutilization when resources can only be scaled at fixed configurations. To facilitate the choice of a suitable MIG configuration for GPU sharing, we further propose a new reward metric that trades off system-level throughput and resource underutilization for different priorities. We use this metric to assess the opportunity of using memory offloading in tested applications. In summary, we made the following contributions in this work:
\begin{itemize}[leftmargin=10pt]
    \item We provide a method for quantifying resource underutilization in different GPU sharing schemes, including MIG, MPS, and time-slicing and apply it to a set of real-world applications from LLM, HPC and data analytics domains;
    \item We identify the opportunities of GPU sharing in improving system throughput and energy, and power throttling as the main interference challenge;
    \item We propose a memory offloading scheme over Nvlink-C2C as a solution to tackle resource underutilization in MIG; %
    \item We define a reward metric that trades off performance and resource underutilization for selecting GPU sharing configurations to meet different targets.
\end{itemize}

\section{Motivation and Background}

\subsection{Single Application GPU Sharing}

In CUDA on Nvidia GPUs, submitting work to the GPU is done through CUDA \textit{stream}~\cite{cuda_runtime_api}. A stream is an ordered queue of commands, e.g., data movement commands, synchronization commands, and GPU kernel launch commands. Commands submitted to two distinct streams might execute concurrently on the GPU, enabling some level of concurrency within a single application that uses multiple streams.

\textit{Green contexts} are lightweight CUDA contexts that can be configured to only execute kernels on a fixed set of the GPU's SMs~\cite{cuda_driver_api}, offering more granular control over CUDA streams. By creating two or more green contexts, applications can control the mapping of GPU kernels to SMs~\cite{shi2025nexus}. For example, an application can create two green contexts, one with 16~SMs and another one with 32~SMs, and execute two kernels concurrently on distinct sets of SMs. %

\subsection{Multi-Application GPU Sharing}
This type of GPU sharing aims at providing several processes with concurrent access to a single physical GPU. Two main GPU sharing schemes can be identified. First \textit{temporal sharing}, where the GPU only executes work submitted by one process at a time, switching between processes to provide the illusion of the processes simultaneously executing work on the GPU. The second scheme, \textit{spatial sharing}, distributed GPU compute resources to concurrently executes computational work from several processes. At the system level, spatial sharing can be achieved through \textit{Multi-Process Service}~(MPS), which partition compute resources, but keeps GPU memory resources and L2 cache shared between processes; and Multi-Instance GPU~(MIG), which partitions both compute and memory resources.

\subsubsection{Time-slicing}
By default, each process on a GPU-accelerated system maintains its own GPU context. The default scheduling strategy, referred to a \textit{time-slicing}, schedules GPU kernels submitted by distinct processes to the GPU in a round-robin fashion. When a process whishes to submit a GPU kernel to the GPU, it is given a fixed amount of time to execute on the whole GPU. %
The context switching between processes comes with a significant performance cost, as no computational work is executed on the GPU during the context switch. With time-slicing, GPU kernels from distinct processes never execute simultaneously on the GPU, however, other operations, such as data copy operations might execute concurrently.

\subsubsection{Multi-Process Service (MPS)}
Multi-Process Service~(MPS)~\cite{nvidia_mps} is a software mechanism to provide spatial sharing. One process, the MPS \textit{server}, maintains a GPU context and regular CUDA applications submit work to the GPU through the server, effectively sharing one GPU context across independent processes. MPS executes GPU kernels submitted by each client on a portion of the GPU streaming multiprocessors~(SM), effectively achieving spatial sharing of the GPU's compute resources. The amount of SMs allocated to each process is controlled by the user as a percentage of the total SM count. %

On Volta GPUs and newer, MPS provides memory segmentation between processes. However, the memory capacity, bandwidth, and L2 cache are always shared between processes and thus, memory interference may occur. %
Moreover, MPS does not provide error isolation. When a GPU kernel in one MPS process generates a fatal GPU fault, all other processes' GPU kernels executing at the same time on the GPU also return with an error. This specificity makes MPS unsuitable for multi-user GPU sharing. %

\subsubsection{Multi-instance GPU}
\begin{figure}[bt]
    \centering
    \includegraphics[width=\linewidth]{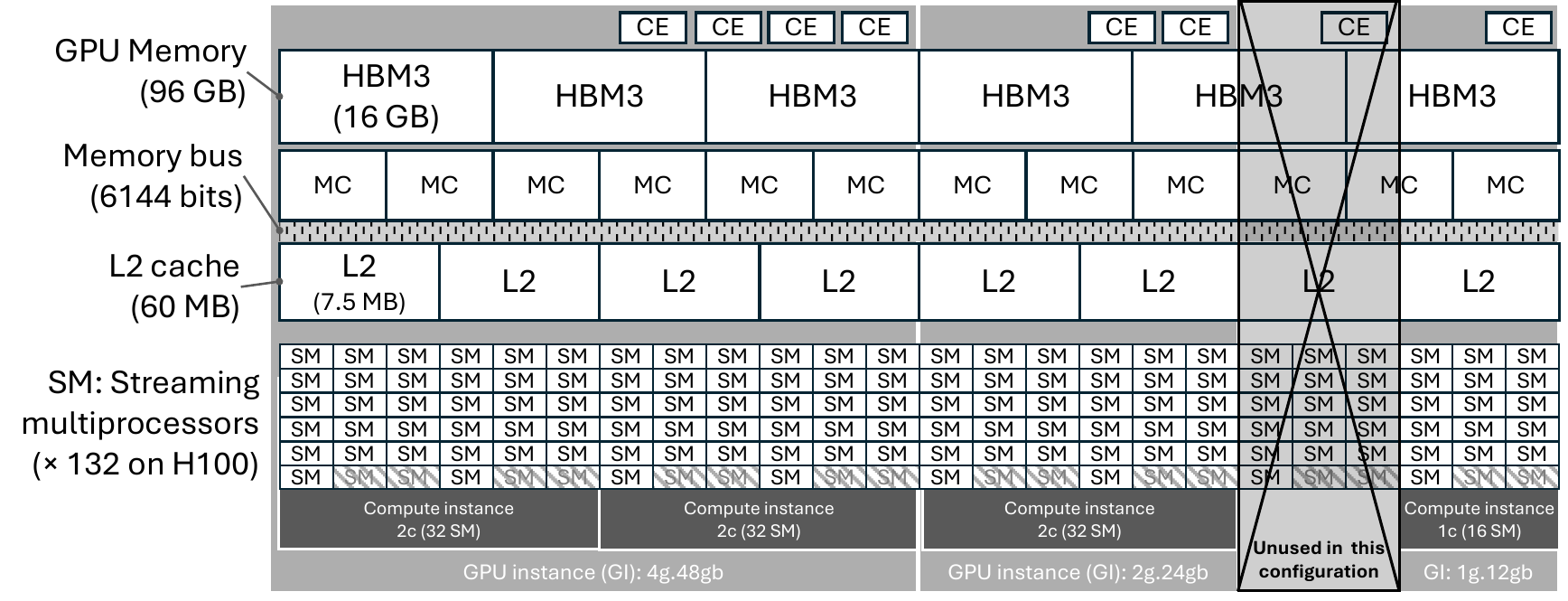}
    \caption{Example MIG configuration on a 96~GB H100 GPU.}
    \label{fig:mig-arch}
\end{figure}

Multi-instance GPU~(MIG)~\cite{nvidia_mig} is a hardware mechanism which partitions a single GPU into multiple smaller \textit{instances}. An instance comprises of a portion of the GPU's compute resources and a portion of the GPU's memory resources. Each instance is exposed to the user as an independent CUDA device, which is selected using the environment variable \verb|CUDA_VISIBLE_DEVICES|.

In MIG, a \textit{GPU instance}~(GI) represents the combination of a portion of GPU's compute resources, a portion of memory resources, and a portion of the copy and decoding engines. This resources division is rigid. Both memory and compute resources are allocated as one or multiple \textit{slices}, with each slice having a fixed size. Each GPU instance gets a fixed number of slices of each type, with a constrained range of possible combinations. %

A memory slice represents roughly one eighth of the total memory resources, which not only comprises of HBM memory, but also a portion of the L2 cache, a number of the GPU's copy engines (CE), and dedicated path across the memory controllers. %
A compute slice represent roughly one seventh of the total compute resources, that is, one seventh of the total SMs count. In reality, the exact SM count significantly deviates from one seventh. Table~\ref{tab:mig} lists the available GPU instance profile for a H100 GPU with 132~SMs. The profile name reflects the number of compute and memory slice for each profile. For example, a \textit{3g.24gb} profile contains \textit{3} compute slices, that is, roughly 2~sevenths of the total SM count, along with 2~eighths of the total 96~GB GPU memory, that is 24~GB.

Once a GPU instance has been created, \textit{compute instances} must be created on top of it for the resources to be usable to the user. A compute instance represents the top-level MIG instance on which users can execute their application. Figure~\ref{fig:mig-arch} presents an example MIG configuration for a 96~GB Nvdia H100 GPU.
MIG allows to create more than one compute instance atop a single GPU instance. In this case, the compute instances share the memory capacity and L2 cache, but each gets its own set of SMs. This approach is similar to MPS. %

Compute instances profiles are named by prefixing the name of the underlying GPU instance with the number of compute slices in the compute instance. For example, \textit{2c.3g.48gb} designates a compute instance with 2 compute slices (``2c''), on top of a \textit{3g.48gb} GPU instance. When the compute instance uses all compute slices of the GPU instance, the first part is omitted: \textit{3c.3g.48gb} is abbreviated \textit{3g.48gb}.

The static configuration in MIG is a known limitation, as changing the MIG configuration is not possible while GPU applications are running. Various solutions have been proposed to tackle this issue, such as dynamic re-configuration~\cite{li2022miso, wu2023transparent}. %

{}

\subsection{Nvlink-C2C Interconnect}
\label{sec:bg-c2c}

Nvlink-C2C interconnect was introduced on the Grace Hopper superchip, where CPU and GPU are connected with it to form a cache-coherent ``chip-to-chip'' interconnect. This high-performance interconnect can deliver up to 450~GB/s bandwidth in each direction. Over Nvlink-C2C, each processor (CPU or GPU) can directly access the other processor's memory at cacheline granularity, namely 128~bytes on the GPU side, and 64~bytes on the CPU side. Cache-coherency is enforced at the hardware level. When memory is accessed over Nvlink-C2C, the data is transparently cached in the cache hierarchy of the accessing processor. Atomics over Nvlink-C2C are also supported. These features are implemented at the hardware level and do not require any user intervention.

\section{Methodology}
\label{sec:setup}

We conduct our experiments on an Nvidia H100 GPU integrated in a Grace Hopper system. This GPU is equipped with 96~GB of HBM3 memory, and features 132~SMs, from the Hopper microarchitecture. Table~\ref{tab:mig} presents the various MIG GPU instance profiles available on this device. It is worth noting here that other H100 GPUs, e.g. with different form factors or memory capacities, might have different characteristics for each supported MIG profile~\cite{nvidia_mig}. In terms of software, we use CUDA runtime~12.4, Nvidia GPU driver version~550.54.15, and Linux kernel~5.14.0. The CPU is a 72-core Arm Neoverse-V2 CPU, equipped with 512~GB of LPDDR5X memory. %

\begin{table}
\caption{A summary of MIG profiles for Grace Hopper H100 (96GB) GPU. Usable resources are given for one MIG instance, wasted resources are GPU-wide best case.}
\label{tab:mig}
\begin{adjustbox}{width=\linewidth,center}{
\centering
\setlength{\tabcolsep}{3pt}
\begin{tabular}{|c|c|c|c|c|c|c|c|c|c|}%
 \hline
 & \textbf{Max.} & \multicolumn{2}{c|}{\textbf{SMs}} & \multicolumn{3}{c|}{\textbf{Memory (GiB)}} & \textbf{L2}&& \textbf{Mem. BW}  \\
\textbf{Profile}  & \textbf{Inst.} & \textbf{Usable} &\textbf{Wasted} &\textbf{Usable} & \textbf{Wasted} &\textbf{\%GPU} & \textbf{Cache} &\textbf{CEs} &\textbf{(GiB/s)} \\\hline\hline
1g.12gb &7  & 16  & 15\%   & 11   & 17.5   &1/8 &1/8 &1 & 406 \\\hline
1g.24gb &4  & 26  & 21\%   & 23   & 2.5   &2/8 &2/8 &2 & 812 \\\hline
2g.24gb &3  & 32  & 3\%    & 23   & 2.5    &2/8 &2/8 &2 & 812  \\\hline
3g.48gb &2  & 60  & 6/9\%  & 46.5 & 1.5  &4/8 &4/8 &3 & 1611 \\\hline
4g.48gb &1  & 64  & 3\%    & 46.5 & 1.5  &4/8 &4/8   &4 & 1635 \\\hline
7g.96gb &1  & 132 & 0\%    & 94.5 & 0 &8/8 &8/8   &8 & 3175 \\\hline
\end{tabular}

}\end{adjustbox}
\end{table}

\subsection{GPU Utilization Metrics}

To quantify GPU utilization, various approach have been proposed, relying on Nvidia's system-level monitoring tool: nvidia-smi to monitor memory usage, Nvidia Nsight System to obtain information on SM utilization, and Nvidia Nsight Compute to obtain information on memory bandwidth utilization, and compute throughput for individual kernels~\cite{strati2024orion,weaver2024granularity}. However, these approaches have significant limitations on pre-Hopper GPUs, as bandwidth utilization and compute throughput can only be measured for individual kernels through overhead-inducing profiling, making system-wide monitoring impractical.
With the introduction of the Hopper microarchitecture, support for sampling of performance metrics has been added to Nvidia Management Library (NVML), under the name GPU Performance Metrics~(GPM). This feature allows sampling GPU performance metrics, either at the GPU level or at the MIG instance level. GPM sampling is exposed to the user in nvidia-smi. However, as this interface provides limited control on the sampling period, we directly use the NVML library to collect GPM metrics.

In detail, we collect several GPM metrics, including those associated with the metrics proposed in~\cite{strati2024orion,weaver2024granularity}. We use a sampling period of 0.2~s. First, we collect the SM utilization metric, which reports the percentage of time that SM are busy in the sampling period. In addition, we collect the more granular SM occupancy metrics, which measures the percentage of active warps relative to the hardware maximum over the sampling period. Finally, we collect the pipeline utilization for arithmetic units and Tensor Cores to identify the behavior of each application in an application-agnostic way. Those metrics are reported for each datatype: double-, single-, and half-/mixed-precision. For memory utilization, we collect both capacity usage and bandwidth usage. In addition to GPM metrics, we also capture regular NVML metrics, namely used GPU memory, GPU power draw, GPU clock frequency, and GPU throttling information. For GPU power draw, NVML API support pooling an history of measurements, with a sampling period of 20~ms, other metrics using the same pooling mechanism have a sampling frequency of 200~ms.

\subsection{Applications}

A key aspect for choosing to run an application under MIG is its low scale, and limited memory footprint. For our study we choose a set of application from various domains: high-performance computing, quantum computing, artificial intelligence, and data analytics. In order to evaluate each application under all available MIG profiles, we select problem sizes that fits in GPU memory for the smallest MIG profile on our system, that is, 12~GB. Table~\ref{tab:workloads} details the selected workloads, and associated data input. 

The \textit{used pipelines} column indicates which of the GPU compute pipelines are used by each application, this information is collected through the pipeline utilization metrics in NVML GPM. Collectively, our selection of workloads utilizes all available compute pipelines on Hopper GPU, with the exception of double precision Tensor Cores, which remain unutilized for all evaluated applications.

In addition to the set of real-world application, we include two variants of the STREAM benchmark~\cite{McCalpin1995}. The STREAM-GPU benchmark executes the STREAM benchmark on the GPU, with data placed locally on the same GPU, this allows us to measure memory bandwidth in MIG. The STREAM-Nvlink variants executes a GPU kernel which reads an array of data placed in CPU memory, and writes back data to another CPU-resident array. This benchmark aims at simulating a heavy load on the CPU-GPU interconnect, in order to evaluate how the Nvlink-C2C link is utilized in MIG, depending on the MIG configuration, and the number of concurrently running applications, on separate MIG instances.

For three of the selected applications, namely FAISS, LLama3, and Qiskit, we also decline a problem version above 12~GB. These problem variants are indicated in parenthesis in Table~\ref{tab:workloads}, and are only used in Section~\ref{sec:offloading}. FAISS is an indexing library~\cite{douze2024faiss}. The core of the benchmark used in this work queries a vector database using approximate nearest neighbor search. In this context, using an index type with a higher memory footprint can speed up query time or improve accuracy. We therefore swap the index type, yielding an overall higher memory footprint when querying the index. It is worth noting that in this variant, the memory footprint of the application only exceeds GPU memory for a very short amount of time, representing a workload with very short memory usage burst. For the LLM inference benchmark, we use the 8 bit-quantized version (``Q8'') of the Llama3-8B model. This quantization allows executing the model on the smallest MIG profile, at the cost of reduced quality. For the larger test case, we use the original half-precision version of the model, increasing inference quality at the cost of higher memory footprint. For Qiskit, the main data structure is the state vector, consisting of $2^n$ single-precision complex numbers, with $n$ the number of simulated qubits. While a 30-qubit state vector fits within 8~GB of GPU memory, a 31-qubit state vector takes up 16~GB of GPU memory, and is therefore unable to fit within the smallest MIG profile memory. We use the 31-qubit case for our larger test case.

For other applications, we do not derive a high-memory variant. AutoDock-GPU does not have a high memory footprint even with large molecule inputs, and therefore does not directly benefit from increasing the GPU memory capacity. NekRS and LAMMPS are codes where the problem to be solved can be user-tuned to fit within given GPU memory. The version of GPT2 training is too limited in scope to exceed the smallest MIG profile memory capacity.

\begin{table}
\caption{A summary of applications. Parenthesis for the input problem indicate a memory footprint above 12~GB.}
\label{tab:workloads}
\begin{adjustbox}{width=\linewidth,center}{
\centering
\begin{tabular}{|l|l|c|c|}
 \hline
 \textbf{Name} &\textbf{Description}  &\textbf{Used Pipelines} &\textbf{Input}\\\hline\hline

Qiskit~\cite{javadi2024quantum} & Quantum Simulation & FP32 & Quantum Volume, 30(-31) qubits \\\hline
FAISS~\cite{douze2024faiss} & Data analytics & FP32, FP16  & sift1M IVF4096,PQ64(/IVF16384)
\\\hline
NekRS~\cite{fischer2022nekrs}   &CFD  & FP64, FP32    & turbPipePeriodic \\\hline
LAMMPS~\cite{thompson2022lammps}  &Molecular Dynamics & FP64 & ReaxFF \\\hline
AutoDock-GPU~\cite{santos2021accelerating}        & Molecular Docking & FP32 & PDBID: 3er5, 2vaa \\\hline
LLM Training~\cite{llm.c}    & GPT2 & HMMA, FP32
   & tinystories, shakespeare~\cite{llmdotc} \\\hline
LLM Inference~\cite{llama_cpp}   & Llama3 & \makecell{HMMA, IMMA,\\ FP32, FP16}
 & Llama~3~8B Q8 (FP16)~\cite{dubey2024llama} \\\hline
Rodinia: hotspot~\cite{che2009rodinia}        & Diff. eq. solver & FP64, FP32 & 1024$\times$1024, 1M iterations \\\hline
STREAM-Nvlink  & Memory Bandwidth & FP64 & 512~MB array\\\hline
STREAM-GPU  & Memory Bandwidth & FP64 & 512~MB array \\\hline

\end{tabular}
}\end{adjustbox}
\end{table}

\subsection{Achievable Resource Utilization in MIG}
\label{sec:achievable}
As a first step, we measure precisely the amount of resources allocated for each MIG profile. While the documentation details that a memory slice is approximately one eighth of the total GPU memory, and a compute slice is approximately one seventh of the GPU's SMs, exact values are not provided.

To measure the number of SMs allocated to one MIG instance, we design a benchmark around a simple GPU kernel. The kernel executes a fixed number of arithmetic operations, in a predictable time. The kernel launch configuration is chosen so that one block fully utilizes one streaming multiprocessor for a fixed amount of time, by setting the block size to the maximum admissible number of threads per block (\verb|maxThreadsPerBlock|). We first launch this kernel with $1$ thread block, measuring the kernel runtime $\Delta t$. We then repeatedly launch the kernel, each time incrementing the thread block count $n$, measuring the runtime for each execution. The smallest $n$ for which the runtime is $2\Delta t$ is such that $n = N_{SM}+1$, with $N_{SM}$ the number of available streaming multiprocessors. This is based on the design of the kernel: when $n$ is equal to the number of streaming multiprocessor, all thread blocks can be scheduled concurrently to run and finish simultaneously after $\Delta t$, with no thread block waiting. Adding one more thread block will result in this block waiting $\Delta t$ for the $N_{SM}$ first blocks to finish before one streaming multiprocessor is freed. We then compare the experimentally-measured streaming multiprocessor count with the number provided by nvidia-smi. In our experiments, those two values matched in all situations. Those values are reported in Table~\ref{tab:mig}.

Using the measured SM counts, we find that when configured with 1g.12gb profile, only 16 SMs can be used in each instance. Since the system can support seven instances of this profile, it means that 15\% SMs ($7\times16=112$ SMs compared 132 total) cannot be utilized on the GPU. We present the percentage of wasted SMs for each profile in Table~\ref{tab:mig}. We note here that this resource waste is due to the limitation of seven GPU instances in MIG. On Ampere A100 GPU, for which MIG was first introduced, this number (7) matches the seven Graphics Processing Clusters~(GPC). On Hopper H100 GPUs, the limitation of seven GPU instances remains, even though the number of GPC is eight.

In addition, we observe that most MIG configurations leave some memory capacity unused. This is listed as wasted memory in Table~\ref{tab:mig}. For example, creating seven MIG 1g.12gb GPU instances will result in 17.5~GiB unused memory capacity. Overall, memory waste is limited compared to SM waste.

\begin{table}[bt]
    \centering
    \captionsetup[table]{skip=0pt}
    \renewcommand{\arraystretch}{1.25}
    \caption{Measured Nvlink-C2C bandwidth using cudaMemcpy and direct in-kernel access (GiB/s).}

\begin{subtable}{\linewidth}
\centering  
\resizebox{0.9\linewidth}{!}{%
    \begin{tabular}{lm{0.75cm}m{0.75cm}m{0.75cm}m{0.75cm}rr}
    
          &               &              &              &                  & Local          & Ratio \\
    Profile & \textbf{BOTH} & \textbf{D2H} & \textbf{H2D} & \textbf{Local} &  \% of total bw & D2H/H2D \\ 
    1g.12gb    & \cellcolor{red!60}41.7          & \cellcolor{red!60}39.6          & \cellcolor{red!60}44.0          & \cellcolor{green!10!yellow!80}357.5              & 13\%           & 0.900    \\ 
    2g.24gb    & \cellcolor{yellow!80}79.2           & \cellcolor{red!60}39.6         & \cellcolor{red!60}44.0           & \cellcolor{green!20!yellow!80}720.5              & 26\%           & 0.900    \\ 
    3g.24gb    & \cellcolor{yellow!80}79.2          & \cellcolor{red!60}39.6          & \cellcolor{red!60}44.0          & \cellcolor{green!50!yellow!80}1383.3             & 50\%           & 0.900    \\
    4g.48gb    & \cellcolor{yellow!80}79.1          & \cellcolor{red!60}39.6         & \cellcolor{red!60}44.0          & \cellcolor{green!50!yellow!80}1438.9             & 52\%           & 0.900    \\ 
    7g.96gb    & \cellcolor{yellow!80}79.2           & \cellcolor{red!60}39.6         & \cellcolor{red!60}44.0          & \cellcolor{green!80!}2732.4             & 100\%          & 0.900   \\
    No MIG& \cellcolor{green!10!yellow!80}329.1           & \cellcolor{green!10!yellow!80}276.3          & \cellcolor{green!10!yellow!80}333.1          & \cellcolor{green!80!}2741.4             & 100\%          & 0.829   \\ 

    \end{tabular}  
}
\caption{cudaMemcpy}
\label{tab:nvlink_ce}
\end{subtable}
    
\begin{subtable}{\linewidth}
\centering  
\resizebox{0.9\linewidth}{!}{%
    \begin{tabular}{lm{0.75cm}m{0.75cm}m{0.75cm}m{0.75cm}rr}
    
          &               &              &              &   & Local          & Ratio \\
    Profile & \textbf{BOTH} & \textbf{D2H} & \textbf{H2D} & \textbf{Local}  &  \% of total bw & D2H/H2D \\ 
    1g.12gb    & \cellcolor{red!40!yellow!80}282           & \cellcolor{yellow!80}343          & \cellcolor{red!60}207          & \cellcolor{yellow!80}406              & 13\%           & 1.65    \\ 
    2g.24gb   & \cellcolor{red!20!yellow!80}334           & \cellcolor{yellow!80}338          & \cellcolor{red!30!yellow!80}303          & \cellcolor{green!20!yellow!80}812              & 26\%           & 1.12    \\ 
    3g.48gb    & \cellcolor{red!20!yellow!80}331           & \cellcolor{yellow!80}336          & \cellcolor{yellow!80}348          & \cellcolor{green!50!yellow!80}1611             & 51\%           & 0.97    \\
    4g.48gb    & \cellcolor{red!20!yellow!80}330           & \cellcolor{yellow!80}338          & \cellcolor{yellow!80}347          & \cellcolor{green!50!yellow!80}1635             & 51\%           & 0.97    \\ 
    7g.96gb    & \cellcolor{red!20!yellow!80}327           & \cellcolor{yellow!80}336          & \cellcolor{yellow!80}341          & \cellcolor{green!80!}3175             & 100\%          & 0.98   \\
    No MIG& \cellcolor{red!20!yellow!80}325           & \cellcolor{yellow!80}336          & \cellcolor{yellow!80}337          & \cellcolor{green!80!}3175             & 100\%          & 1.00    \\ 
    \end{tabular}
}
\caption{Direct GPU access}
\label{tab:nvlink_direct}
\end{subtable}

\removespacehere
\end{table}

\subsection{Nvlink-C2C}
MIG distributes compute and memory resources across GPU instances, two distinct GPU instances do not compete for memory bandwidth, or compute time. However, the Nvlink-C2C CPU-GPU interconnect is shared between MIG instances, which raises the question of understanding how Nvlink-C2C can be utilized by an application running under MIG. 

Within GPU applications, we identify two possible approaches to transfer data over Nvlink-C2C between CPU and GPU memory. The first approach relies on calling explicit data movement APIs, such as cudaMemcpy. Those APIs offload data movements to the GPU's copy engines~(CE). As copy engines are distributed across MIG instances, we expect applications using this approach to not interfere with each others when concurrently performing CPU-GPU data movements. However, as one instance only gets a fraction of the GPU's CEs, the transfer bandwidth over Nvlink-C2C is also limited to a fraction of the overall bandwidth, even if only one MIG instance is performing data transfer. The second approach relies on direct access ability of the GPU's streaming multiprocessors to access CPU-resident memory directly from within GPU kernels.

We design a benchmark to characterize the achievable bandwidth for each of those two approaches. The benchmark performs a copy operation from a source to a destination buffer. By controlling the respective placements of source and destination buffers, we evaluate host-to-device, device-to-host, and local memory bandwidth. For example, host-to-device is obtained by placing source buffer in host (CPU) memory, and destination buffer in device (GPU) memory. Finally, by executing two copy operations in opposite directions in parallel, we measure the bidirectional bandwidth. We derive two variants of this benchmark. The first variant uses a GPU copy kernel to evaluate performance of direct access, inspired by the STREAM benchmark~\cite{McCalpin1995}. The second variant uses cudaMemcpy to perform the copy operation, evaluating performance of copy engine-performed data movements.

Results for the cudaMemcpy benchmark are presented in Table~\ref{tab:nvlink_ce}. The smallest MIG instance, 1g.12gb, achieves respectively 13\% and 14\% for H2D and D2H data movements, relative to the Nvlink-C2C bandwidth measured with MIG disabled. As a 1g MIG instance gets one of the eight GPU's copy engines, these results are expected. However, increasing the MIG instance size does not provide bandwidth improvement, even though more compute engines are advertised for larger MIG instances. This issue is likely a bug, and might get fixed in future driver/runtime versions.

Table~\ref{tab:nvlink_direct} presents the measured direct access bandwidth for each of the MIG profile. For the local memory bandwidth, the achieved bandwidth is also presented as a percentage of the total memory bandwidth, measured with MIG disabled. We observe that this percentage matches the ratio of total memory allocated to each profile, which is consistent with the observation that memory bandwidth is split across MIG instances. We also observe that even for the smallest MIG profile, the direct access benchmark is able to saturate the Nvlink-C2C interconnect in device-to-host direction. This is a key observation, as direct access enable an application running on the smallest MIG instance is able fully utilize the Nvlink-C2C bandwidth, which is not possible with a cudaMemcpy operation.

Overall, for any MIG profile, direct GPU access provides the highest bandwidth over Nvlink-C2C, up to 338~GB/s in device-to-host direction, and 348~GB/s in host-to-device direction. However, since this approach relies on a GPU kernel to perform the copy, it cannot be overlapped with computations.

\section{Imbalanced Resource Underutilization}
In this section, we analyze resource utilization of compute and memory, respectively, when applications running on full GPUs or GPU sharing. We also characterize performance-resource scaling in applications. 

\subsection{Compute Resource utilization}

\label{sec:compute-util}
\begin{figure}[bt]
    \centering
    \includegraphics[width=\linewidth]{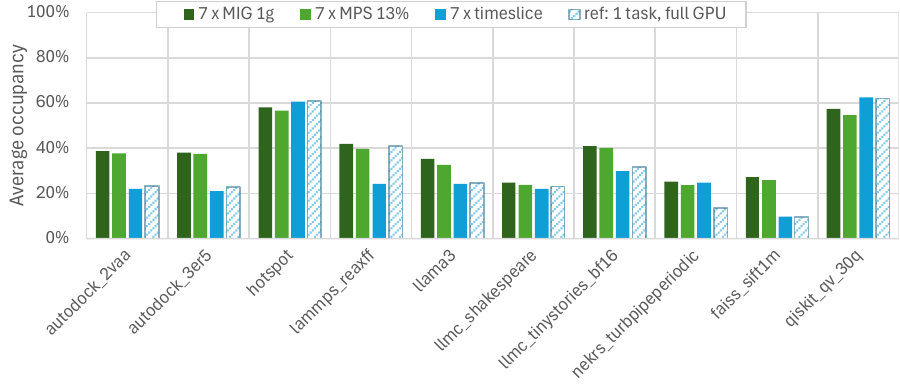}
    \vspace{-17pt}
    \caption{GPU compute resource utilization, measured as the SM occupancy using different GPU sharing options.}
    \label{fig:sm}
\end{figure}

To quantify the compute resources utilization, we measure the average achieved SM occupancy over the entire execution of each workload. Three GPU sharing mechanisms are compared, namely MPS, MIG, and time slice, and the full GPU execution is used for the reference. The results are presented in Figure~\ref{fig:sm}. \textit{Streaming Multiprocessor}~(SM) is the building block of Nvidia GPUs. Each SM can execute instructions independently, and a GPU consists of multiple SMs. At a high level, the number of SMs directly correlates to the compute capacity of a GPU. As shown in Table~\ref{tab:mem_compute_gpus}, the last four generations of GPUs have not only increased the compute throughput per SM but also packed more SMs in a GPU. If a workload has low SM occupancy, indicating there are not enough active warps to be scheduled to SMs, the peak hardware compute capacity cannot be fully exploited. %

For workloads that already achieve high SM occupancy in the full GPU execution, GPU sharing may cause their SM occupancy to decrease. For instance, Qiskit, LAMMPS, and hotspot achieve 60\%, 40\%, and 60\% SM occupancy on a full GPU. When running them in the three GPU sharing configurations, their SM occupancy either slightly decreases in MIG and MPS, as in Qiskit and hotspot, or even sharply drops by half in time slicing in LAMMPS. In contrast, if a workload is under-utilizing compute resources in full GPU execution, running in GPU sharing may bring substantial improvement in compute resource utilization. One such example is NekRS, whose SM occupancy is nearly doubled from 12\% to 25\% in all three GPU sharing configurations. Similarly, the two AutoDock variants have their SM occupancy nearly doubled in GPU sharing.

The compute resources underutilization observed at the application level can be explained by two distinct causes at the GPU kernel level. First, as GPU compute resources are increased, kernel execution time on the GPU diminishes. If the shortened GPU execution time is unmatched by a decrease in the CPU execution time, the GPU might become idle. One such case is NekRS, where the CPU-side execution dominates and keeps the GPU idle. The second cause may come from the tail effect in SM scheduling. GPU kernels are divided into thread blocks, which are scheduled onto SMs. Some blocks might finish earlier than others. If no block is available to schedule, some SMs become idle. In the worst case of high imbalance, only one SM remains active while all others are idle. On larger GPUs with more SMs, this tail effect leads to more SMs being left idle compared to smaller GPUs. Such tail effect is observed in the AutoDock workloads.

Among the three GPU sharing options, time slicing results in the lowest SM occupancy in most cases, except for hotspot and Qiskit. For instance, AutoDock-3er5 using time-slicing only achieves 20\% SM occupancy, while it achieves $38-39$\% SM occupancy in MIG and MPS, nearly a 1.9$\times$ improvement. The high context switch cost in time slicing could be the cause for the overall reduced occupancy. For hotspot and Qiskit, since they have already achieved high SM occupancy in full GPU execution, their lower SM occupancy in GPU sharing could be caused by the wasted SMs (SMs that cannot be scheduled) as explained in Section~\ref{sec:achievable}. MPS always underperforms by 1-5\% compared to MIG in all test cases. This could be caused by memory interference, as MPS does not impose any memory isolation. Overall, workloads with low SM utilization can leverage MIG to reduce compute resource underutilization while incurring lower overhead than other options, such as MPS and time slicing.

\subsection{Memory resource utilization}
\begin{figure}[bt]
    \centering
    \includegraphics[width=\linewidth]{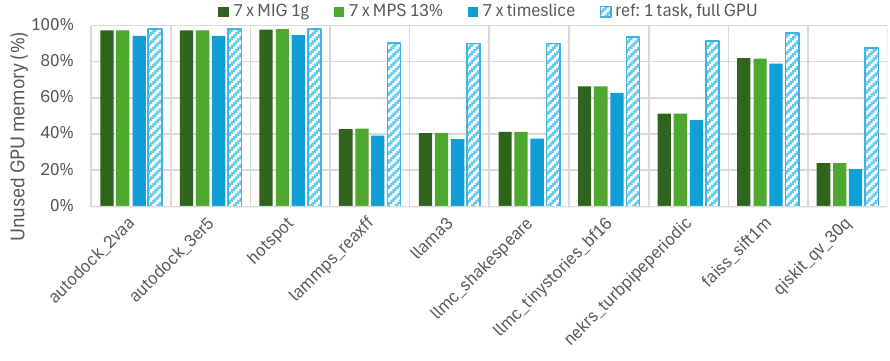}
    \includegraphics[width=\linewidth]{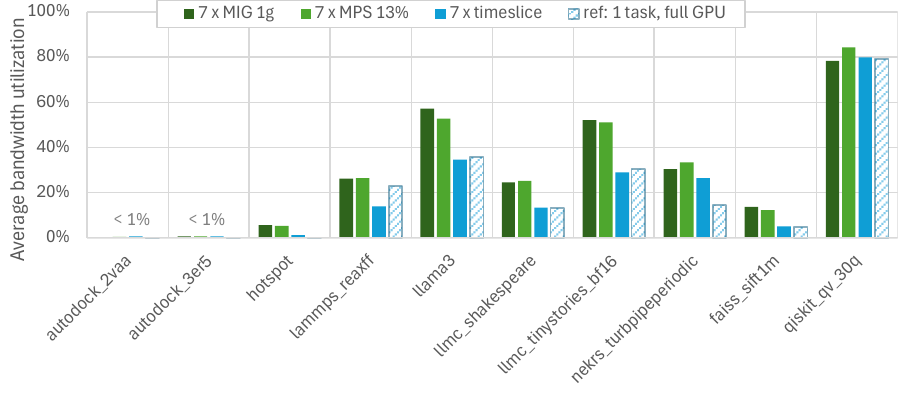}
    \vspace{-17pt}
    \caption{GPU memory capacity (upper) and bandwidth (lower) utilization.}
    \label{fig:mem}
    \removespacehere
\end{figure}

We analyze memory resource utilization in two aspects, i.e., memory capacity and memory bandwidth. Both aspects directly affect the cost of a GPU system and performance, and thus their utilization needs to be evaluated separately. The memory capacity is often perceived as a hard requirement, as insufficient memory capacity prevents running applications with higher memory needs. On the other hand, bandwidth resource is a more performance-critical requirement, especially for workloads that can efficiently utilize the computing power of GPUs, such as LLM training, whose performance highly depends on memory bandwidth.

The top panel of Figure~\ref{fig:mem} presents the memory capacity underutilization, computed as the used memory relative to total GPU memory, reported by nvidia-smi. As expected, in most cases, using MIG slices significantly reduces unused GPU memory capacity. In fact, seven cases see reduced memory underutilization, by $20$-$60$\%. AutoDock and hotspot have very low memory footprints and are therefore still significantly underutilizing GPU memory, even with GPU sharing. Memory capacity underutilization arises because the workloads execute with a fixed maximum memory footprint, and do not make use of GPU memory above this footprint. For example, the 8~billion parameters half-precision LLama~3 model fits within 16~GB of memory. This also holds true within a range of scientific applications with fixed problem size, e.g., the AutoDock workloads can fit within 1~GB of GPU memory. %

We find that time slicing has lower memory waste compared to MIG and MPS. However, since this is a system-level measurement, the memory usage might also include GPU context-induced overhead. To quantify this overhead, we implement a benchmark that initializes a CUDA context without running any computation, allocating memory resources necessary to execute a CUDA application; this is achieved using \verb|cudaMalloc(NULL)|. We measure the memory overhead as $\sim$60MB per process in MIG 1g.12gb and $\sim$600MB per process in time slicing. For MPS, we measure a fixed total overhead of $\sim$600~MB, independent of the number of processes. Clearly, time slicing has the highest GPU context-induced memory overhead, which explains why memory underutilization at system level is lower in time slicing, although the memory is not used by workloads.

The bottom panel of Figure~\ref{fig:mem} presents the memory bandwidth utilization. Only half of the workloads can effectively utilize the memory bandwidth. In these cases, using GPU sharing, as compared to full GPU execution, results in improved bandwidth utilization in all cases, except LAMMPS on time slicing, which has reduced bandwidth utilization. Among the three GPU sharing schemes, time slicing has significantly lower bandwidth utilization, e.g., llama3 achieves 35\% utilization in time slicing but 58\% in MIG and 52\% in MPS. The reduced bandwidth in time slicing is caused by the high context switch overhead, during which no memory operations are performed, reducing the average bandwidth utilization over the entire execution. NekRS and Qiskit achieve higher memory bandwidth utilization in MPS compared to MIG. However, as MPS does not isolate cache between applications, increased memory bandwidth utilization might indicate increased cache conflicts between co-running tasks, generating heavier memory accesses. We identified this to be the case here, as NekRS and Qiskit present degraded performance when co-running under MIG.

\subsection{Coarse-grained Performance-Resource Scaling}
\label{sec:scaling}
\begin{figure}[bt]
    \centering
    \includegraphics[width=\linewidth]{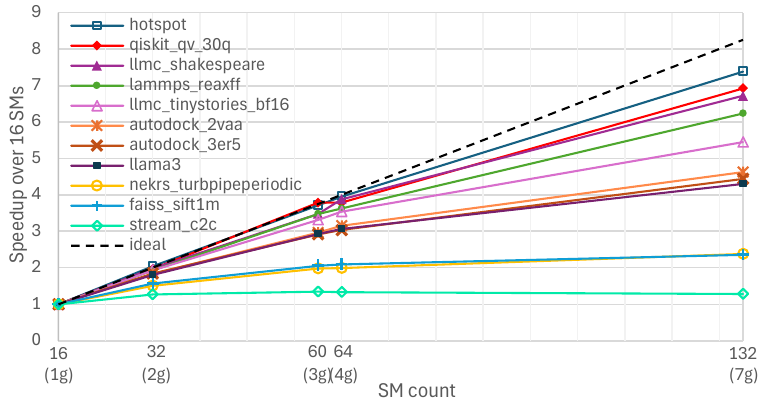}
    \caption{GPU Performance-Resource Scaling for each application on a Nvidia Grace Hopper system.}
    \label{fig:scaling}
\end{figure}

From the SM occupancy and memory utilization in previous sections, we find that an application may have imbalanced usage of compute and memory resources. For instance, hotspot has high utilization of compute resources but low memory usage. Qiskit has up to 60\% SM occupancy but nearly 90\% memory bandwidth usage. In this experiment, we investigate another cause for resource underutilization, i.e., the imbalanced performance scaling as the amount of resources increase. Current resource provisioning via MIG slice is coarse-grained, where GPU compute and memory resources are coupled and increase at nearly a doubling rate. This often mismatches the imbalanced usage of compute and memory resources in real applications. We conduct a performance-resource scaling test, as reported in Figure~\ref{fig:scaling}, where each workload runs with increased resources as provisioned with increased MIG slice from 1g.12gb up to 7g.96gb. We report their relative performance by normalizing to the performance obtained in the smallest MIG 1g profile.

As the problems in each workload remain constant, with the increased resources, not all workloads scale up their performance proportionally as indicated by the dashed line in Figure~\ref{fig:scaling}. Qiskit and hotspot scale the closest to the ideal scaling, with LAMMPS and llmc following a similar scaling. They represent the first category of workloads that have good performance-resource scalability, likely because they achieve balanced high utilization of both compute and memory. The second class of workloads, including AutoDock and llama3 inference, shows a certain deviation from the ideal scaling. Finally, the class of workloads exhibiting the worst scaling is NekRS, FAISS, and STREAM, which heavily utilize memory resources but have limited compute resources.

A further investigation indicates that resource underutilization in fixed MIG slices may increase rapidly. This may severely waste resources, especially for the second and third classes of workloads, whose performance-resource scalability is less than ideal. For instance, memory underutilization grows at a higher than doubling rate when increasing the MIG slice. Therefore, the resource requirements of applications scale in a way that mismatches the coarse-grained MIG profiles. There is a need to combine resource underutilization and performance into consideration when selecting suitable GPU sharing options.

\section{Opportunities and Challenges in GPU sharing}
In this section, we analyze opportunities in system-level throughput and energy consumption when GPU sharing is enabled. We also identify power throttling interference as a potential challenge.

\subsection{Co-running System Throughput}
\begin{figure}
    \centering
    \includegraphics[width=\linewidth]{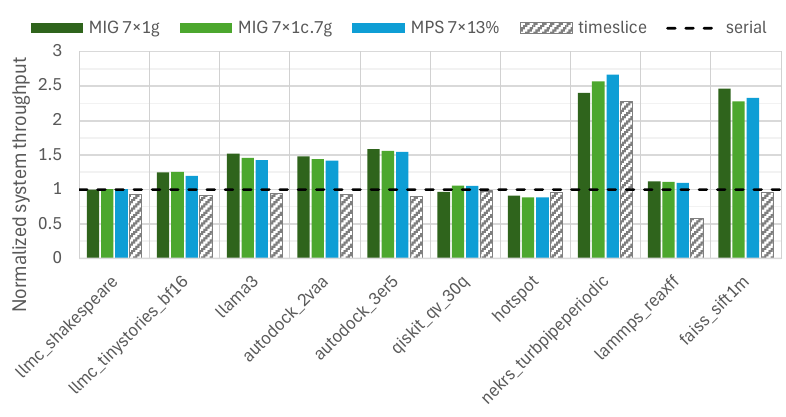}
    \vspace{-15pt}
    \caption{System throughput for concurrent execution of seven workload copies, normalized to serial execution baseline.}
    \vspace{-10pt}
    \label{fig:systemthroughput}
\end{figure}

\begin{figure}
    \centering
    \includegraphics[width=\linewidth]{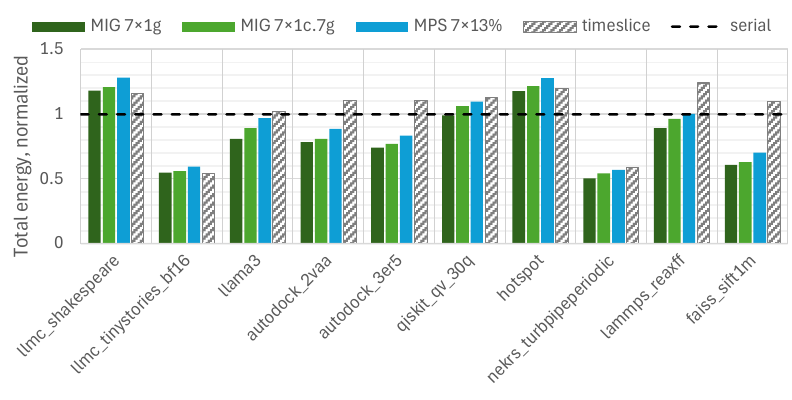}
    \vspace{-15pt}
    \caption{Total energy consumed for concurrently running 7 GPU-sharing tasks, normalized to the total energy consumed for seven serial runs of the same application.}
    \label{fig:energy}
\end{figure}

We measured aggregated task throughput when running seven instances of an application concurrently across different GPU sharing schemes.
Figure~\ref{fig:systemthroughput} shows normalized system throughput, with serial execution as a baseline, obtained by sequentially running seven times the same application on the entire GPU. Overall, running seven applications in parallel on GPU sharing configurations achieves an average throughput improvement of approximately 1.4$\times$ compared to sequential execution, demonstrating the benefit of concurrent execution for system-level efficiency.

We measured aggregated task throughput when running seven instances of an application concurrently across different GPU sharing schemes. Figure~\ref{fig:systemthroughput} shows normalized system throughput, where values above 1 indicate improvement over serial execution of the same seven instances. Overall, running seven applications in parallel on GPU sharing configurations achieves an average throughput improvement of approximately 1.4$\times$ compared to sequential execution, demonstrating the benefit of concurrent execution for system-level efficiency.

Two applications achieve exceptional throughput gains: NekRS and FAISS reach 2.4$\times$ and 2.5$\times$ improvement with GPU sharing with MIG 7$\times$1g respectively. This aligns with their resource utilization characteristics. Both exhibit low GPU compute utilization (13.5\% and 10\% SM occupancy respectively in Figure~\ref{fig:sm}), substantial unused GPU memory (over 50\% as shown in Figure~\ref{fig:mem}), and relatively low bandwidth utilization. Running on smaller MIG instances results in better occupancy and memory utilization, as discussed in Section~\ref{sec:compute-util}, increasing overall system throughput.%

Conversely, compute-intensive workloads show minimal benefit or slight degradation. Qiskit and hotspot experience less than 5\% throughput reduction, consistent with their high compute resource utilization (62\% and 61\% SM occupancy respectively in Figure~\ref{fig:sm}). %

Comparing sharing mechanisms, MIG 7$\times$1g generally achieves higher throughput than MIG 7$\times$1c.7g and MPS 7$\times$13\%, except for Qiskit and NekRS. These two applications achieve higher memory bandwidth utilization under MPS than MIG (as shown in Figure 3), since because MIG's strict memory isolation limits each partition to one-seventh of the total bandwidth, while MPS allows more flexible bandwidth sharing. For bandwidth-sensitive workloads, this hardware-enforced isolation in MIG can become a performance bottleneck despite providing better resource guarantees.

\subsection{Co-running Energy Utilization}

To quantify the energy implications of GPU sharing with MIG, we measure total energy consumption for different sharing configurations. The instantaneous GPU-level power consumption is sampled every 20ms using NVML and these measurements are integrated over the total duration to compute total energy. We compared running seven concurrent identical tasks under different GPU sharing configurations against serial execution baseline. Figure~\ref{fig:energy} shows the normalized results. The results reveal compelling energy efficiency benefits across nearly all workloads tested. 

GPU sharing demonstrates substantial energy savings in most applications, with reductions ranging from 10\% (LAMMPS) variants to over 50\% (NekRS). On average, workloads achieve 26\% energy reduction compared to serial execution. These savings stem from two primary sources: the elimination of GPU idle periods between sequential task executions, and more efficient amortization of static power consumption across multiple concurrent workloads. However, llmc\_shakespeare and hotspot exhibit increased energy consumption under GPU sharing, while Qiskit shows comparable energy to serial execution, suggesting workload-specific sensitivity to sharing mechanisms.

Among the four evaluated GPU sharing configurations, MIG 7$\times$1g consistently achieves the lowest energy consumption across all benchmarks, on average reducing total energy to 63\% of the serial baseline. This fine-grained partitioning approach outperforms MIG 7$\times$1c.7g, MPS 7$\times$13\%, and timeslice scheduling in every workload tested. The efficiency of MIG 7×1g suggests that maximizing parallelism through smaller partitions optimally balances resource utilization with power consumption, providing the most compelling energy efficiency for datacenter deployments where operational costs and environmental impact are critical considerations.

\subsubsection{Power Throttling Interference}
\begin{figure}[tb]
    \centering
    \begin{subfigure}{\linewidth}
        \centering
        \includegraphics[width=0.49\linewidth]{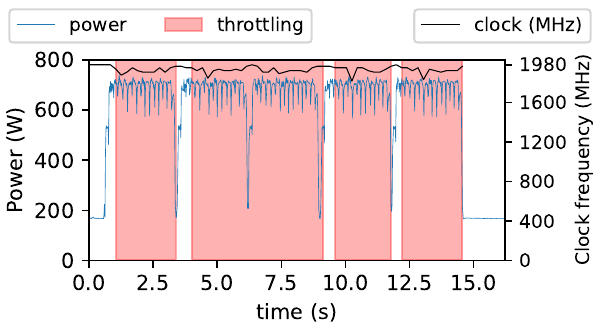}
        \includegraphics[width=0.49\linewidth]{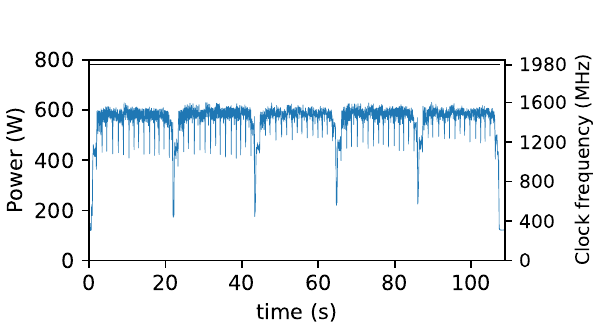}
        \vspace{-17pt}
        \caption{Qiskit}
        \vspace{10pt}
        \label{fig:powerandthrottling-qiskit}
    \end{subfigure}
    \begin{subfigure}{\linewidth}
        \centering
        \includegraphics[width=0.49\linewidth]{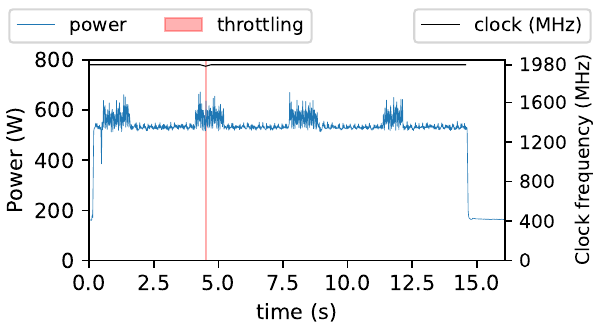}
        \includegraphics[width=0.49\linewidth]{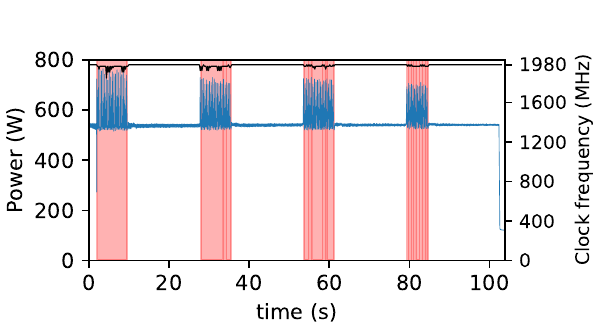}
        \vspace{-17pt}
        \caption{LLM training}
        \vspace{10pt}
        \label{fig:powerandthrottling-llama}
    \end{subfigure}
    \caption{Power consumption and throttling behavior for (a) Qiskit and (b) LLM training. Left: single application on full GPU. Right:  concurrent executions on 7$\times$1g MIG partitions. Pink regions indicate power throttling periods.}
    \label{fig:powerandthrottling}
\end{figure}

By monitoring GPU power draw and clock frequencies at 20ms intervals during concurrent execution, we observe power throttling behavior. Figure~\ref{fig:powerandthrottling} shows representative power traces for memory-bound (Qiskit) and compute-intensive (LLM training) workloads under both isolated (left) and concurrent (right) execution scenarios.

Memory-bound workloads experience reduced power consumption under GPU sharing. Qiskit running on the full GPU (Figure~\ref{fig:powerandthrottling-qiskit}, left) exhibits continuous throttling, hitting the 700W power limit and dropping clock frequencies from 1980~MHz down to 1815~MHz. When seven instances of Qiskit execute in parallel, each on its 1g.12gb MIG instance (Figure~\ref{fig:powerandthrottling-qiskit}, right), we observe a maximum power consumption of 670W, below the 700W power cap. Therefore, no throttling is observed. The MIG bandwidth partitioning naturally constrains power draw below throttling thresholds, eliminating throttling but reducing per-instance performance.

Compute-intensive workloads trigger power throttling only during concurrent execution. LLM training running alone (Figure~\ref{fig:powerandthrottling-llama}, left) maintains 500-650~W power consumption without throttling. However, seven concurrent instances on independent MIG 1g instances (Figure ~\ref{fig:powerandthrottling-llama}, right) collectively exceed the 700~W limit, causing periodic throttling with associated frequency drop. This occurs despite performance isolation promises in MIG, as hardware partitioning isolates compute and memory resources but not power delivery.

Above observations reveal a potential challenge in GPU sharing: while MIG provides logical resource isolation, shared power delivery creates a side channel for interference.

\section{GPU Sharing with Nvlink-C2C enabled Offloading}
\label{sec:offloading}

\subsection{Nvlink-C2C Offloading}

The characteristics of MIG instances are fixed, and the choice of a MIG profile is limited to a fixed set of possible memory capacities, that are multiples of the smallest memory slice, namely 12~GB on our platform. When a workload's memory footprint is slightly above the memory capacity of a given MIG profile, the next larger profile must be used to enable execution of the workload. Such choice comes with a significant increase in memory capacity, even though the workload to execute might only utilize a fraction of this added memory capacity. Taking the example of Qiskit, a 31-qubit simulation consumes 16~GB of GPU memory, that is, 4~GB more than the smallest 12~GB MIG profile. However, running such workload requires selecting a 24~GB MIG profile, leaving 8~GB of memory unused, that is 33\% of the GPU instance memory.

For such workloads, instead of increasing the MIG instance size, we propose to offload part of the application's data to host (CPU) memory. During GPU computations, part of the data is accessed over Nvlink-C2C. We implement such offloading strategy for three of the applications, as described in Section~\ref{sec:setup}, namely FAISS, Qiskit, and Llama3. We implement the offloading scheme in FAISS and Llama3 by using a memory allocator that supports spilling out of GPU memory, namely \verb|cudaMallocManaged| or \verb|malloc|, focusing on performing this change on the large data structures used by each application. For FAISS this is the index data, and for Llama3, we change the allocator used in the GGML backend. We try a similar implementation in Qiskit. However, the natively-supported memory swapping strategy used by Qiskit to spill the state vector over GPU memory appeared to consistently outperform our implementation, we therefore use this swapping strategy to serve the purpose of offloading part of the application's data to host memory.

\subsection{Reward Model}
The offloading scheme is expected to increase resource utilization. In Section~\ref{sec:scaling} we detail how, for a given application, SM occupancy tends to increase as the instance size is reduced. Therefore, our offloading scheme is expected to enable increasing the SM occupancy due to the reduced MIG instance size. In addition, the offloading scheme reduces the unused GPU memory. However, data access over Nvlink-C2C comes with a significant performance cost, an application executing with our offloading scheme might therefore suffer significant performance degradation.

To quantify the tradeoff between application performance and resource utilization, we design a simple parametrized cost-benefit model. For this purpose, we consider a MIG instance with a memory capacity ${M}_\text{instance}$ and $N_\text{SM}$ streaming multiprocessors. We execute the application under study on this MIG instance, measuring the application performance $P$, the average occupancy at the GPU level ${Occ}$, and the peak memory usage $M_\text{app}$. By normalizing the occupancy to the MIG instance, based on the total number of SM on the GPU $N_\text{SM,GPU}$, we define the \textit{SM waste}, noted $W_\text{SM}$. Analogously, we define the \textit{memory waste}, $W_\text{MEM}$:

\begin{equation*}
\begin{cases}
W_\text{SM} = \frac{N_\text{SM}}{N_\text{SM,GPU}}(1-Occ)  \vspace{0.1cm} \\
W_\text{MEM} = \frac{M_\text{instance}-M_\text{app}}{M_\text{GPU}}
\end{cases}
\end{equation*}

We measure the application performance $P_\text{GPU}$, obtained when running the application on the entire GPU. From the SM waste, the memory waste, the relative performance $P/P_\text{GPU}$, and a dimensionless term $\alpha$ we derive the following cost-benefit model:

\begin{equation*}
R = \frac{P/P_{\text{GPU}}}{\alpha + W_\text{MEM} + W_\text{SM}} 
\end{equation*}

For a fixed application and a fixed value of $\alpha$, we aim at finding the MIG configuration that maximizes $R$. The term $\alpha$ represents a policy used to establish whether the model should prioritize reducing resource underutilization ($\alpha=0$), or to prioritize increasing performance, by setting a positive value for $\alpha$. As $\alpha$ is increased above 0, the model shifts from a utilization-only model to a performance-only model. Since both $W_{SM}$ and $W_{MEM}$ are within $[0,1]$, we vary $\alpha$ within the same range, to ensure it operates on a comparable scale. Increasing $\alpha$ beyond $1$ this range would further favor performance at the cost of greater resource underutilization.%

\subsection{Results}
\begin{figure}
    \centering

    \includegraphics[width=0.9\linewidth]{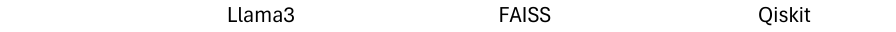}
    \includegraphics[width=0.9\linewidth]{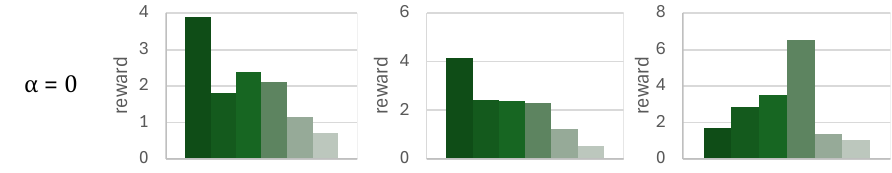}\hrule
    \includegraphics[width=0.9\linewidth]{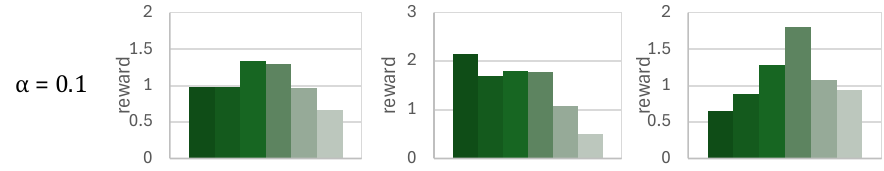}\hrule
    \includegraphics[width=0.9\linewidth]{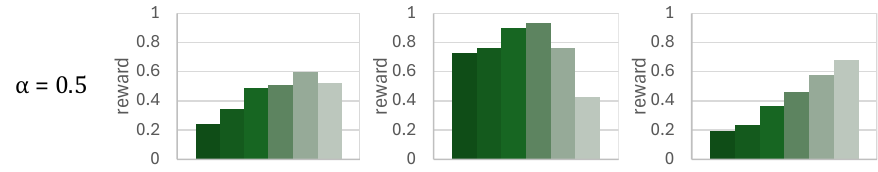}\hrule
    \includegraphics[width=0.9\linewidth]{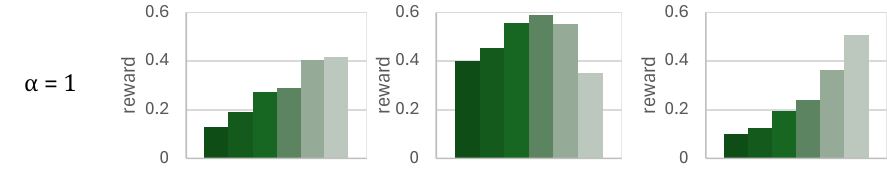}
    \includegraphics[width=\linewidth]{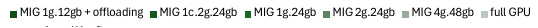}
    
    \caption{Reward based selection ($\alpha=0, 0.1, 0.5, 1$) in three applications.}
    \vspace{-17pt}
    \label{fig:reward0}
\end{figure}

We apply our model to the three selected applications. For Qiskit and FAISS, we define the performance metric $P$ as the inverse of the runtime. For Llama3, the performance $P$ is the number of token generated per second during the text generation phase of the benchmark. Higher value of $P$ indicates higher performance. Figure~\ref{fig:reward0} presents the reward value produced by our model for a set of MIG configurations for three different applications, with different policies $\alpha$. The results when using the proposed offloading scheme is also presented, with a MIG instance size of 1g.12gb.

Looking at the first row, with $\alpha=0$, that is, we prioritize reducing resource waste, without penalizing low performance. With this choice, the offloading scheme yields the highest reward for both FAISS and Llama3. However, for Qiskit, the most favored configuration for $\alpha=0$ is 2g.24gb. This is because in this application, the occupancy is the highest for the 2g.24gb case, making this case the best from the perspective of reducing resource underutilization.

When increasing $\alpha$ to $0.1$, the offloading scheme is only preferred for the FAISS case. We suggest that this is because FAISS offloads a small fraction of the overall application working set to host memory, for a very short period of time, resulting in limited performance degradation. Therefore, when including performance to a limited extent in our model ($\alpha>0$), the offloading configuration is still preferred.

Setting $\alpha$ to $1$ shifts the preference towards MIG instances with larger sizes, which is expected as this variant of the model prioritizes performance over reducing underutilization. For Llama3 and Qiskit, the largest configuration, namely full GPU, is preferred. This is expected, as discussed in Section~\ref{sec:scaling}, those two workload, exhibit a near-perfect performance scaling behavior as the number of streaming multiprocessor is increased. In contrast, for FAISS, our model indicates a preference for the 2g.24gb configuration. This is because this particular test case exhibits poor scaling behavior when increasing MIG size, as discussed in Section~\ref{sec:scaling}.

\section{Related Works}

Characterization of GPU sharing techniques have previously been conducted. Gilman et al.~\cite{gilman2022characterizing} compare priority streams, time-slicing and MPS. For MPS, Xu et al.~\cite{xu2022igniter} quantify interference risk by measuring L2 cache misses. Architectural and software changes have also been proposed to improve multi-application concurrency~\cite{ausavarungnirun2018mask,li2024star,pavlidakis2024guardian}. Li et al~\cite{li2022characterizing} provides an evaluation of MIG performance and energy consumption for machine learning workloads. Works tend to demonstrate a reduced interference risk compared to MPS~\cite{li2022miso,zhang2024miger}. While existing works focus on a single range of workload, notably deep learning, we instead offer an application-agnostic approach, covering a broad scope of workloads spanning scientific computing, AI, and data analytics.

Existing works on MIG have focused on mitigating impact of co-locating workloads on GPUs at the scheduler level. Tan et al.~\cite{tan2021serving} propose an approach to identify optimal MIG partitioning for DNN model serving. MIG has also been considered at the cluster level for LLM model serving~\cite{turkkan2024optimal,zhu2024optimizing}. Combined use of MPS and MIG have also been proposed; either by using measured MPS performance to execution with MIG~\cite{li2022miso}, or by using MPS atop MIG for execution~\cite{zhang2024miger}. Our work identifies GPU resource underutilization issues, and propose a cost-benefit model to evaluate the trade-off between performance and resource utilization across MIG configurations.

Memory offloading to CPU memory has been proposed as a solution to overcome GPU memory limitations. Studies of this approach over Nvlink-C2C have been conducted~\cite{werner2025towards,schieffer2024harnessing}. We use memory offloading as a solution to reduce resource underutilization, caused by the rigidity of MIG configurations.

\section{Conclusion}
This work provides a first in-depth exploration of using GPU sharing to tame resource underutilization in HPC applications, including real-world scientific, LLM, and data analytics applications. We propose multiple metrics to quantify compute and memory resource underutilization, respectively. Our results indicate that the coarse-grained resource scaling of tightly coupled compute and memory resources, as provided in MIG slices, mismatches the resource-performance scaling characteristics of applications. To address the discontinuity in resource provisioning due to fixed MIG profiles, we leverage Nvlink-C2C offloading to enable workloads slightly larger than a MIG slice to run without requesting a larger MIG profile. Moreover, we propose a new reward metric to trade off the system-level throughput and resource underutilization, and evaluated different choices of GPU sharing in eight workloads using the reward metric on a Grace Hopper system.

\section*{Acknowledgments}
This research is supported by the Swedish Research Council (no. 2022.03062). This work has received funding from the European High Performance Computing Joint Undertaking (JU) and Sweden, Finland, Germany, Greece, France, Slovenia, Spain, and the Czech Republic under grant agreement No.~101093261.

\bibliographystyle{IEEEtran}
\bibliography{main}

\begin{thebibliography}{10}
\providecommand{\url}[1]{#1}
\csname url@samestyle\endcsname
\providecommand{\newblock}{\relax}
\providecommand{\bibinfo}[2]{#2}
\providecommand{\BIBentrySTDinterwordspacing}{\spaceskip=0pt\relax}
\providecommand{\BIBentryALTinterwordstretchfactor}{4}
\providecommand{\BIBentryALTinterwordspacing}{\spaceskip=\fontdimen2\font plus
\BIBentryALTinterwordstretchfactor\fontdimen3\font minus \fontdimen4\font\relax}
\providecommand{\BIBforeignlanguage}[2]{{%
\expandafter\ifx\csname l@#1\endcsname\relax
\typeout{** WARNING: IEEEtran.bst: No hyphenation pattern has been}%
\typeout{** loaded for the language `#1'. Using the pattern for}%
\typeout{** the default language instead.}%
\else
\language=\csname l@#1\endcsname
\fi
#2}}
\providecommand{\BIBdecl}{\relax}
\BIBdecl

\bibitem{han2022microsecond}
M.~Han, H.~Zhang, R.~Chen, and H.~Chen, ``Microsecond-scale preemption for concurrent {GPU-accelerated} {DNN} inferences,'' in \emph{16th USENIX Symposium on Operating Systems Design and Implementation (OSDI 22)}, 2022, pp. 539--558.

\bibitem{strati2024orion}
F.~Strati, X.~Ma, and A.~Klimovic, ``Orion: Interference-aware, fine-grained gpu sharing for ml applications,'' in \emph{Proceedings of the Nineteenth European Conference on Computer Systems}, 2024, pp. 1075--1092.

\bibitem{gregg2012fine}
C.~Gregg, J.~Dorn, K.~Hazelwood, and K.~Skadron, ``{Fine-Grained} resource sharing for concurrent {GPGPU} kernels,'' in \emph{4th USENIX Workshop on Hot Topics in Parallelism (HotPar 12)}, 2012.

\bibitem{li2022miso}
B.~Li, T.~Patel, S.~Samsi, V.~Gadepally, and D.~Tiwari, ``Miso: exploiting multi-instance gpu capability on multi-tenant gpu clusters,'' in \emph{Proceedings of the 13th Symposium on Cloud Computing}, 2022, pp. 173--189.

\bibitem{li2022characterizing}
B.~Li, V.~Gadepally, S.~Samsi, and D.~Tiwari, ``Characterizing multi-instance {GPU} for machine learning workloads,'' in \emph{2022 IEEE International Parallel and Distributed Processing Symposium Workshops (IPDPSW)}.\hskip 1em plus 0.5em minus 0.4em\relax IEEE, 2022, pp. 724--731.

\bibitem{wu2023transparent}
B.~Wu, Z.~Zhang, Z.~Bai, X.~Liu, and X.~Jin, ``Transparent $\{$GPU$\}$ sharing in container clouds for deep learning workloads,'' in \emph{20th USENIX Symposium on Networked Systems Design and Implementation (NSDI 23)}, 2023, pp. 69--85.

\bibitem{weaver2024granularity}
A.~Weaver, K.~Kavi, D.~Milojicic, R.~P.~H. Enriquez, N.~Hogade, A.~Mishra, and G.~Mehta, ``Granularity-and interference-aware gpu sharing with mps,'' in \emph{SC24-W: Workshops of the International Conference for High Performance Computing, Networking, Storage and Analysis}.\hskip 1em plus 0.5em minus 0.4em\relax IEEE, 2024, pp. 1630--1637.

\bibitem{nvidia_mps}
Nvidia, ``Nvidia multi-process service overview (r575),'' 2025.

\bibitem{pavlidakis2024guardian}
M.~Pavlidakis, G.~Vasiliadis, S.~Mavridis, A.~Argyros, A.~Chazapis, and A.~Bilas, ``Guardian: Safe gpu sharing in multi-tenant environments,'' in \emph{Proceedings of the 25th International Middleware Conference}, 2024, pp. 313--326.

\bibitem{shi2025nexus}
X.~Shi, C.~Cai, J.~Du, Z.~Zhu, and Z.~Jia, ``Nexus: Taming throughput-latency tradeoff in llm serving via efficient gpu sharing,'' \emph{arXiv e-prints}, pp. arXiv--2507, 2025.

\bibitem{tan2021serving}
C.~Tan, Z.~Li, J.~Zhang, Y.~Cao, S.~Qi, Z.~Liu, Y.~Zhu, and C.~Guo, ``Serving dnn models with multi-instance gpus: A case of the reconfigurable machine scheduling problem,'' \emph{arXiv preprint arXiv:2109.11067}, 2021.

\bibitem{gilman2022characterizing}
G.~Gilman and R.~J. Walls, ``Characterizing concurrency mechanisms for nvidia gpus under deep learning workloads,'' \emph{ACM SIGMETRICS Performance Evaluation Review}, vol.~49, no.~3, pp. 32--34, 2022.

\bibitem{cuda_runtime_api}
Nvidia, ``Cuda runtime api documentation,'' 2025.

\bibitem{cuda_driver_api}
------, ``Cuda driver api documentation,'' 2025.

\bibitem{nvidia_mig}
------, ``Nvidia multi-instance gpu user guide,'' 2024.

\bibitem{McCalpin1995}
J.~D. McCalpin, ``Memory bandwidth and machine balance in current high performance computers,'' \emph{IEEE Computer Society Technical Committee on Computer Architecture (TCCA) Newsletter}, pp. 19--25, Dec. 1995.

\bibitem{douze2024faiss}
M.~Douze, A.~Guzhva, C.~Deng, J.~Johnson, G.~Szilvasy, P.-E. Mazar{\'e}, M.~Lomeli, L.~Hosseini, and H.~J{\'e}gou, ``The faiss library,'' \emph{arXiv preprint arXiv:2401.08281}, 2024.

\bibitem{javadi2024quantum}
A.~Javadi-Abhari, M.~Treinish, K.~Krsulich, C.~J. Wood, J.~Lishman, J.~Gacon, S.~Martiel, P.~D. Nation, L.~S. Bishop, A.~W. Cross \emph{et~al.}, ``Quantum computing with qiskit,'' \emph{arXiv preprint arXiv:2405.08810}, 2024.

\bibitem{fischer2022nekrs}
P.~Fischer, S.~Kerkemeier, M.~Min, Y.-H. Lan, M.~Phillips, T.~Rathnayake, E.~Merzari, A.~Tomboulides, A.~Karakus, N.~Chalmers \emph{et~al.}, ``Nekrs, a gpu-accelerated spectral element navier--stokes solver,'' \emph{Parallel Computing}, vol. 114, p. 102982, 2022.

\bibitem{thompson2022lammps}
A.~P. Thompson, H.~M. Aktulga, R.~Berger, D.~S. Bolintineanu, W.~M. Brown, P.~S. Crozier, P.~J. In't~Veld, A.~Kohlmeyer, S.~G. Moore, T.~D. Nguyen \emph{et~al.}, ``Lammps-a flexible simulation tool for particle-based materials modeling at the atomic, meso, and continuum scales,'' \emph{Computer physics communications}, vol. 271, p. 108171, 2022.

\bibitem{santos2021accelerating}
D.~Santos-Martins, L.~Solis-Vasquez, A.~F. Tillack, M.~F. Sanner, A.~Koch, and S.~Forli, ``Accelerating autodock4 with gpus and gradient-based local search,'' \emph{Journal of chemical theory and computation}, vol.~17, no.~2, pp. 1060--1073, 2021.

\bibitem{llm.c}
A.~Karpathy, ``llm.c: Llm training in simple, raw c/cuda,'' \url{https://github.com/karpathy/llm.c}, 2024.

\bibitem{llmdotc}
``llm.c,'' \url{https://github.com/karpathy/llm.c}, 2024.

\bibitem{llama_cpp}
G.~Gerganov and contributors, ``llama.cpp: Inference of llama models in pure c/c++,'' \url{https://github.com/ggerganov/llama.cpp}, 2023.

\bibitem{dubey2024llama}
A.~Dubey, A.~Jauhri, A.~Pandey, A.~Kadian, A.~Al-Dahle, A.~Letman, A.~Mathur, A.~Schelten, A.~Yang, A.~Fan \emph{et~al.}, ``The llama 3 herd of models,'' \emph{arXiv e-prints}, pp. arXiv--2407, 2024.

\bibitem{che2009rodinia}
S.~Che, M.~Boyer, J.~Meng, D.~Tarjan, J.~W. Sheaffer, S.-H. Lee, and K.~Skadron, ``Rodinia: A benchmark suite for heterogeneous computing,'' in \emph{2009 IEEE international symposium on workload characterization (IISWC)}.\hskip 1em plus 0.5em minus 0.4em\relax Ieee, 2009, pp. 44--54.

\bibitem{xu2022igniter}
F.~Xu, J.~Xu, J.~Chen, L.~Chen, R.~Shang, Z.~Zhou, and F.~Liu, ``igniter: Interference-aware gpu resource provisioning for predictable dnn inference in the cloud,'' \emph{IEEE Transactions on Parallel and Distributed Systems}, vol.~34, no.~3, pp. 812--827, 2022.

\bibitem{ausavarungnirun2018mask}
R.~Ausavarungnirun, V.~Miller, J.~Landgraf, S.~Ghose, J.~Gandhi, A.~Jog, C.~J. Rossbach, and O.~Mutlu, ``Mask: Redesigning the {GPU} memory hierarchy to support multi-application concurrency,'' \emph{ACM SIGPLAN Notices}, vol.~53, no.~2, pp. 503--518, 2018.

\bibitem{li2024star}
B.~Li, Y.~Wang, T.~Wang, L.~Eeckhout, J.~Yang, A.~Jaleel, and X.~Tang, ``Star: Sub-entry sharing-aware tlb for multi-instance gpu,'' in \emph{2024 57th IEEE/ACM International Symposium on Microarchitecture (MICRO)}.\hskip 1em plus 0.5em minus 0.4em\relax IEEE, 2024, pp. 309--323.

\bibitem{zhang2024miger}
B.~Zhang, S.~Li, and Z.~Li, ``Miger: Integrating multi-instance gpu and multi-process service for deep learning clusters,'' in \emph{Proceedings of the 53rd International Conference on Parallel Processing}, 2024, pp. 504--513.

\bibitem{turkkan2024optimal}
B.~Turkkan, P.~Murali, P.~Harsha, R.~Arora, G.~Vanloo, and C.~Narayanaswami, ``Optimal workload placement on multi-instance gpus,'' \emph{arXiv preprint arXiv:2409.06646}, 2024.

\bibitem{zhu2024optimizing}
Y.~Zhu, C.~Wang, M.~Calman, R.~Nakazawa, and E.~K. Lee, ``Optimizing gpu multiplexing for efficient and cost-effective access to diverse large language models in gpu clusters,'' in \emph{2024 32nd International Conference on Modeling, Analysis and Simulation of Computer and Telecommunication Systems (MASCOTS)}.\hskip 1em plus 0.5em minus 0.4em\relax IEEE, 2024, pp. 1--8.

\bibitem{werner2025towards}
F.~Werner, M.~Weisgut, and T.~Rabl, ``Towards memory disaggregation via nvlink c2c: Benchmarking cpu-requested gpu memory access,'' in \emph{Proceedings of the 4th Workshop on Heterogeneous Composable and Disaggregated Systems}, 2025, pp. 8--14.

\bibitem{schieffer2024harnessing}
G.~Schieffer, J.~Wahlgren, J.~Ren, J.~Faj, and I.~Peng, ``Harnessing integrated cpu-gpu system memory for hpc: a first look into grace hopper,'' in \emph{Proceedings of the 53rd International Conference on Parallel Processing}, 2024, pp. 199--209.

\end{thebibliography}

\end{document}